\documentclass[twocolumn,aps,prc,superscriptaddress,nofootinbib,floatfix,longbibliography]{revtex4-1}
\RequirePackage{lineno}
\usepackage{url}
\usepackage{cancel}
\usepackage[colorlinks,linkcolor=blue,citecolor=blue,filecolor=black,urlcolor=blue]{hyperref}
\usepackage{epsfig,graphics}
\usepackage{graphicx}
\usepackage{dcolumn}
\usepackage{bm}
\usepackage{amssymb}
\usepackage{amsmath}
\usepackage{multirow}
\usepackage{float}
\usepackage{harpoon}
\usepackage{MnSymbol}
\usepackage{appendix}
\usepackage{color}
\usepackage{hyperref}
\usepackage{cleveref}

\newcommand{\nch} {N_{\mathrm{ch}}}

\newcommand{\pT} {p_{\mathrm{T}}}

\newcommand{\lr}[1]{\left\langle #1\right\rangle}
\newcommand{\lrb}[1]{\left[ #1\right]}

\newcommand{\npart}{N_{\mathrm{part}}}

\newcommand{\nqt}{N_{\mathrm{3q}}}
\newcommand{\nqs}{N_{\mathrm{7q}}}

\graphicspath{{./}{data/}}

\begin{document}
\title{Symmetric-asymmetric collision comparison: disentangling nuclear structure and subnucleonic structure effects for small system flow}
\newcommand{\bnl}{Physics Department, Brookhaven National Laboratory, Upton, NY 11976, USA}
\newcommand{\sbu}{Department of Chemistry, Stony Brook University, Stony Brook, NY 11794, USA}
\newcommand{\moe}{Key Laboratory of Nuclear Physics and Ion-beam Application (MOE), and Institute of Modern Physics, Fudan
University, Shanghai 200433, China}
\newcommand{\fudan}{Shanghai Research Center for Theoretical Nuclear Physics, NSFC and Fudan University, Shanghai 200438, China}

\author{Shengli Huang}\affiliation{\sbu}
\author{Jiangyong Jia}\email[Correspond to\ ]{jiangyong.jia@stonybrook.edu}\affiliation{\sbu}\affiliation{\bnl}
\author{Chunjian Zhang}\affiliation{\moe}\affiliation{\fudan}\affiliation{\sbu}

\begin{abstract}
Previous flow measurements in small collision systems were mostly based on highly asymmetric collisions ($p$+Pb, $p$+Au, $d$+Au, $^{3}$He+Au), where both nuclear structure and subnucleonic fluctuations are important. Comparing these asymmetric systems with the newly available symmetric $^{16}$O+$^{16}$O collisions at RHIC and LHC provides a unique opportunity to disentangle these two contributions. Using Glauber models incorporating both nucleon and quark-level substructure, we analyze multiplicity distributions and initial-state estimators: eccentricities $\varepsilon_n$ for anisotropic flow $v_n$ and inverse transverse size $d_{\perp}$ for radial flow. We find that subnucleonic fluctuations impact O+O collisions differently from asymmetric systems, creating specific patterns in flow observables that enable disentangling the competing contributions. Such experimental comparisons will reduce uncertainties in the initial conditions and improve our understanding of the properties of the QGP-like medium produced in small systems.
\end{abstract}
\maketitle

{\bf Introduction.} The discovery of collective flow phenomena in small collision systems has challenged our understanding of quark-gluon plasma (QGP) formation and its dynamical evolution~\cite{Grosse-Oetringhaus:2024bwr,Noronha:2024dtq}. While anisotropic flow $v_n$ and radial flow in large collision systems are well-established as hydrodynamic responses to initial geometric anisotropies, their origin in small systems remains debated~\cite{Busza:2018rrf}. Previous measurements have focused mostly on highly asymmetric collisions: $pp$~\cite{CMS:2010ifv,ATLAS:2015hzw,CMS:2016fnw} and $p$+Pb~\cite{CMS:2012qk,ALICE:2012eyl,ATLAS:2012cix} at the LHC, and $p$+Au, $d$+Au, $^{3}$He+Au at RHIC~\cite{PHENIX:2013ktj,PHENIX:2018lia,STAR:2022pfn,STAR:2023wmd}, where both nuclear structure and subnucleonic fluctuations contribute significantly to the initial geometry. These two sources of geometric fluctuations are deeply entangled in asymmetric systems, making it difficult to fully establish whether the observed flow patterns arise from geometry-driven final-state response or alternative mechanisms such as initial momentum anisotropy~\cite{Huang:2019tgz}.

A key advantage in studying small ion collisions is that nuclear structure inputs are well controlled since their nucleon configurations can reliably modelled through {\it ab initio} approaches with realistic nuclear interactions~\cite{Hergert:2020bxy,Ekstrom:2022yea}. In contrast, the subnucleon degrees of freedom, especially their spatial structure, are mostly based on simple phenomenological approaches~\cite{Mantysaari:2016ykx,Loizides:2016djv}. This gap in our theoretical understanding makes the comparison between small collision systems particularly compelling, as we can vary the nuclear structure while testing the role of subnucleonic contributions.

The recent collection of symmetric $^{16}$O+$^{16}$O collision data at both RHIC~\cite{Huang:2023viw} and the LHC~\cite{oolhc} presents a unique opportunity to resolve this ambiguity. Symmetric systems provide an initial geometry dominated by the average shape of the overlap region, whereas asymmetric systems are dominated by fluctuations while producing similarly sized QGP-like droplets. Therefore, comparison between the two types of systems enables the disentanglement of nuclear structure effects from subnucleonic fluctuations through a direct comparison of flow observables between symmetric and asymmetric systems.

In hydrodynamic models, anisotropic flow coefficients $v_n$ arise from the linear response to initial eccentricities $\varepsilon_n$~\cite{Niemi:2012aj}, while radial flow characterized by mean transverse momentum $[\pT]$ responds to the inverse system size $d_{\perp}$~\cite{Bozek:2012fw}. If small system flow is indeed driven by final-state hydrodynamic response to initial geometry, then the controlled variation between symmetric and asymmetric collision geometries should produce specific, predictable differences in flow observables. Conversely, if initial momentum anisotropy dominates, one would expect the geometric distinctions to be suppressed, with more similar impacts on $v_2$ and $v_3$ at same multiplicities across different collision systems.

Previous studies have shown that subnucleonic fluctuations are necessary to explain the large $v_n$ observed in $pp$ and $p$+Pb collisions at the LHC. At RHIC, the interpretation of the $p/d/^{3}$He+Au results depends critically on assumptions about subnucleonic structure: models considering only nucleon degrees-of-freedom predict $v_{3}^{\mathrm{He+Au}} > v_{3}^{d\mathrm{Au}} \approx v_{3}^{p\mathrm{Au}}$, while including subnucleonic fluctuations yields $v_{3}^{\mathrm{He+Au}} \approx v_{3}^{d\mathrm{Au}} \approx v_{3}^{p\mathrm{Au}}$~\cite{Zhao:2022ugy,Wu:2023vqj}. Though the latter is in agreement with STAR measurements at mid-rapidity~\cite{STAR:2022pfn}, the former is consistent with PHENIX measurements using correlations between mid-rapidity and backward regions~\cite{PHENIX:2018lia}. This tension may arise from longitudinal decorrelations affecting the PHENIX results~\cite{Zhao:2022ugy}, which is not inconsistent with the presence of subnucleonic fluctuations. Nevertheless, the dependence on initial condition modeling and measurement methods has limited our ability to extract medium properties and distinguish between competing mechanisms~\cite{He:2015hfa,Kurkela:2019kip,Lacey:2025bos}.

This paper presents a systematic study of initial-state estimators, $\varepsilon_2$, $\varepsilon_3$, and $d_{\perp}$, for collective flow and multiplicity distributions $p(\nch)$, across $p/d/^{3}$He+Au and O+O collisions. The impact of subnucleonic structure on these estimators is found to differ significantly between symmetric and asymmetric systems, providing clear signatures for disentangling these geometric sources.

{\bf Model Setup.} 
We calculate initial geometry quantities $\varepsilon_n$, $d_{\perp}$, and $\nch$ using a standard Glauber model framework~\cite{Loizides:2016djv} for both asymmetric ($p/d/^{3}$He+Au) and symmetric ($^{16}$O+$^{16}$O) collisions at $\sqrt{s_{_{\mathrm{NN}}}}=200$ GeV with a nucleon-nucleon inelastic cross section of 42 mb. The impact of subnucleonic fluctuations are evaluated with three variants of Glauber models: nucleon Glauber (no substructure), three-quark Glauber, and seven-quark Glauber. 

Nucleons are assumed to have a 0.4 fm hard-core radius. The nucleon configurations within deuteron and $^{3}$He follow Hulthen wavefunction and Green's function MC calculations with AV18+UIX interactions~\cite{Loizides:2014vua}, respectively, while nucleon configurations within $^{16}$O are generated using nuclear lattice effective field theory (NLEFT)~\cite{Giacalone:2024luz}. These approaches provide ab initio descriptions of nuclear structure based on realistic nucleon-nucleon and three-nucleon interactions. Nucleons in Au are sampled from a standard Woods-Saxon density profile~\cite{Loizides:2016djv}.

The subnucleonic fluctuations are modelled via a phenomenological approach using the quark Glauber models. They distribute quarks inside nucleons according to the ``mod'' configuration~\cite{Mitchell:2016jio}, ensuring the recentered radial distribution follows the proton form factor $\rho_{\mathrm{proton}}(r) = e^{-r/r_0}$ with $r_0=0.234$ fm~\cite{DeForest:1966ycn}. The quark-quark inelastic cross-section is set to 8.1 mb and 1.9 mb for three-quark Glauber model and seven-quark Glauber model, respectively.

The eccentricity and inverse size of the nuclear overlap area in each collision event are calculated as:
\begin{align}\label{eq:1}
\varepsilon_n &= |\lrb{(x+iy)^n}|/\lrb{|(x+iy)|^n}\;,\\\label{eq:2}
d_{\perp} &= 1/\sqrt{\lrb{x^2}\lrb{y^2}}\;,
\end{align}
where the coordinates $(x,y)$ are defined relative to the center-of-mass of the participating nucleons or quarks, and $\lrb{..}$ denotes averaging within a single event. From these, we calculate the two- and four-particle cumulants of $\varepsilon_n$, 
\begin{align}\label{eq:3}
\varepsilon_n\{2\} &= \sqrt{\lr{\varepsilon_n^2}}\;,\\\label{eq:4}
\varepsilon_n\{4\} &= \sqrt[4]{2\lr{\varepsilon_n^2}^2-\lr{\varepsilon_n^4}}\;,
\end{align}
as well as the mean and variance of $d_{\perp}$, $\lr{d_{\perp}}$ and $\sigma_{d_{\perp}} = \sqrt{\lr{(\delta d_{\perp})^2}}$ where $\delta d_{\perp}= d_{\perp}-\lr{d_{\perp}}$. The $\lr{..}$ denotes averaging over an event ensemble.

In large collision systems, $p(\varepsilon_2)$ follows approximately a Bessel-Gaussian distribution~\cite{ATLAS:2013xzf}. In this case, $\varepsilon_2\{4\}$ characterizes the average elliptic shape of the overlap region, while $\varepsilon_2\{2\}$ contains both the average geometry and the random fluctuation contribution. Therefore, the fluctuation component can be isolated from the two- and four-particle cumulants~\cite{Jia:2022qgl}:
\begin{align}\label{eq:5}
\delta_{\varepsilon_2}^2 = \varepsilon_2\{2\}^2-\varepsilon_2\{4\}^2\;.
\end{align}
In smaller collision systems, however, $p(\varepsilon_2)$ has significant non-Gaussian contribution. In this case, $\varepsilon_2\{4\}$ can be non-zero even without an average elliptic shape. This caveat should be kept in mind when using Eq.~\ref{eq:5} in small systems.

For each collision event, we obtain the number of sources $N_s$: the number of participating nucleons $\npart$, number of quarks $\nqt$ for three constituent quarks per-nucleon and $\nqs$ for seven constituent quarks per-nucleon, respectively. Table~\ref{tab:1} shows the average number of participating nucleon or quark sources in minimum bias events. Only a fraction of the quarks within collided nucleons participate in the collision, calculated as $F_3 = \lr{\nqt/3}/\lr{\npart}$ and $F_7 = \lr{\nqs/7}/\lr{\npart}$ for three-quark and seven-quark Glauber models, respectively. $F_3$ and $F_7$ represent the effective fractional mass of the participating nucleons involved in the collisions. This fraction is smallest in $p$+Au and largest in O+O collisions, with an increase of about 20--25\%.

\begin{table}[htbh]
 \begin{tabular}{c|c|cc|cc}
  & Nucleon & \multicolumn{2}{c|}{Three-Quark} & \multicolumn{2}{c}{Seven-Quark}  \\
  & Glauber & \multicolumn{2}{c|}{Glauber} & \multicolumn{2}{c}{Glauber} \\\hline
  & $\lr{\npart}$  & $\lr{\nqt}$ &$F_3$ & $\lr{\nqs}$ & $F_7$ \\\hline
$p$+Au      & 5.94 & 9.29  & 0.521 & 13.24 & 0.318 \\
$d$+Au      & 8.73 & 14.36 & 0.548 & 20.89 & 0.342\\
$^{3}$He+Au & 11.23& 19.53 & 0.580 & 29.04 & 0.369\\
O+O         & 9.72 & 18.27 & 0.627 & 27.06 &0.398\\\hline
\end{tabular}
\caption{The mean values of number of participating nucleons or quarks in minimum bias $p$+Au, $d$+Au, $^{3}$He+Au and $^{16}$O+$^{16}$O collisions obtained from Glauber models including nucleon or subnucleon fluctuations~\cite{Loizides:2016djv}. The effective fraction of participating nucleons in three-quark Glauber model $F_3$ and seven-quark Glauber model $F_7$ are also provided.}
\label{tab:1}
\end{table}

\begin{figure*}[htbp]
\centering
\includegraphics[width=0.85\linewidth]{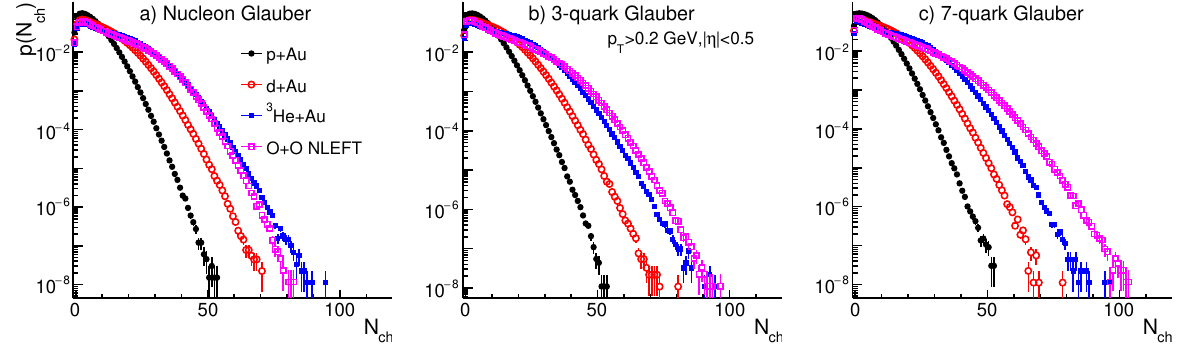}
\caption{Charged particle multiplicity distributions $p(\nch)$ in $p$+Au, $d$+Au, $^{3}$He+Au and $^{16}$O+$^{16}$O collisions at $\sqrt{s_{\mathrm{NN}}}$=200~GeV calculated using nucleon Glauber (left), three-quark Glauber (middle), and seven-quark Glauber (right) models. The enhanced sensitivity of symmetric O+O collisions to subnucleonic fluctuations is evident in the significantly broadened high-multiplicity tails when quark substructure is included, while asymmetric systems show minimal changes across the three models.}
\label{fig:1}
\end{figure*}
\begin{figure*}[htbp]
\centering
\includegraphics[width=0.85\linewidth]{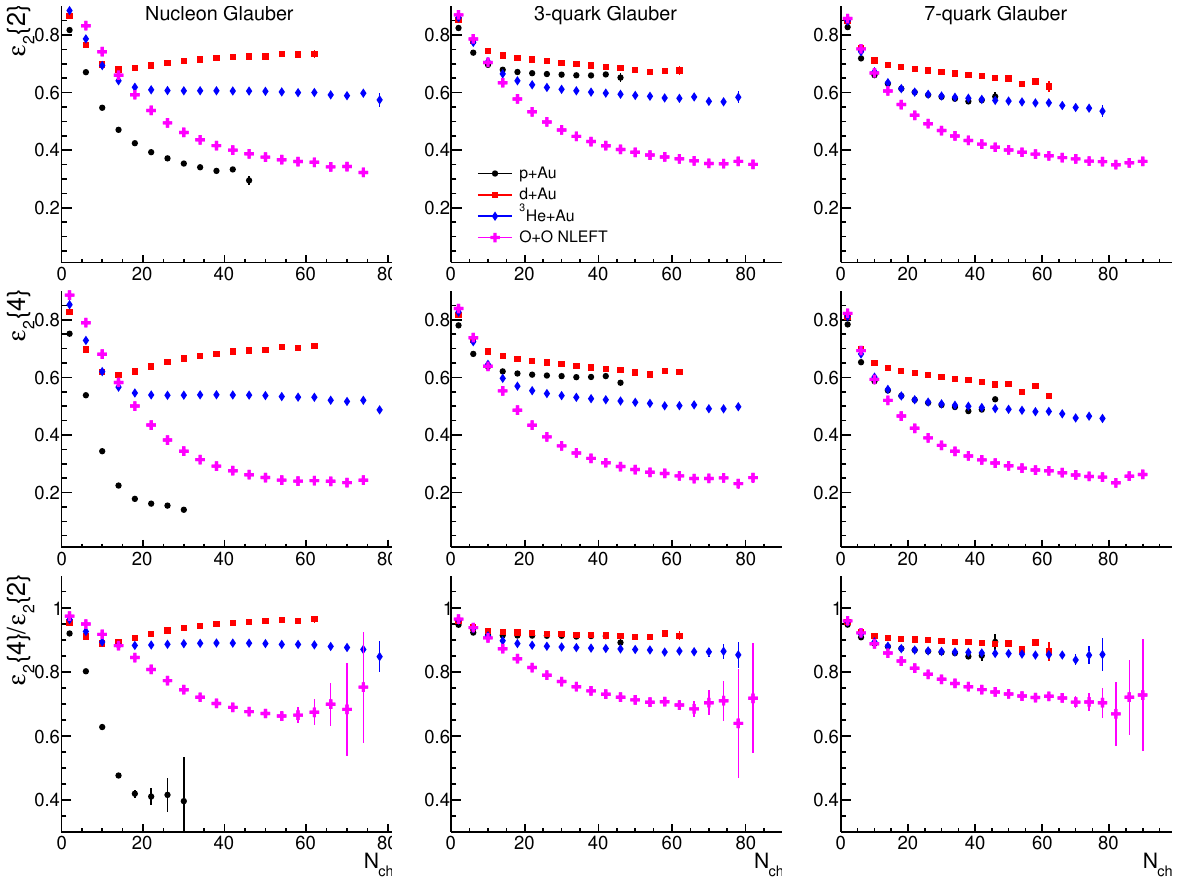}
\caption{Elliptic eccentricity as a function of $\nch$ in $p$+Au, $d$+Au, $^{3}$He+Au, and $^{16}$O+$^{16}$O collisions. Results are shown for nucleon Glauber (left), three-quark (middle), and seven-quark (right) Glauber models. From top to bottom: $\varepsilon_{2}\{2\}$, $\varepsilon_{2}\{4\}$, and $\varepsilon_{2}\{2\}/\varepsilon_{2}\{4\}$.}
\label{fig:2}
\end{figure*}

We assume the particle production for each nucleon or quark source to follow a negative binomial distribution:
\begin{eqnarray}
\label{eq:10}
p(n) = \frac{(n+m-1)!}{(m-1)!n!} p^n(1-p)^m,\; p = \frac{\bar{n}}{\bar{n}+m}\;,
\end{eqnarray}
where $\bar{n}$ controls the average multiplicity of each source, while $\bar{n}$ and $m$ together control its relative fluctuation,
\begin{eqnarray}
\label{eq:11}
\hat{\sigma}^2 \equiv \frac{\left\langle(n-\bar{n})^2\right\rangle}{\bar{n}^2}=\frac{1}{\bar{n}}+\frac{1}{m}\;.
\end{eqnarray}

The total multiplicity $\nch$ is obtained by summing $n$ over all sources in the event. The parameter values for the three-quark Glauber model are obtained by fitting the experimental multiplicity data for $\pT>0.2$ GeV and $|\eta|<0.5$ in Ru+Ru collisions~\cite{STAR:2021mii}; they are $\bar{n}=0.65346$ and $m=0.75148$. For the nucleon Glauber and the seven-quark Glauber models, $\bar{n}$ and $m$ values are rescaled by $1/F_3$ and $F_7/F_3$ such that $\lr{\nch}$ in the minimum bias events are the same across the three Glauber models.

\begin{figure*}[htbp]
\centering
\includegraphics[width=0.85\linewidth]{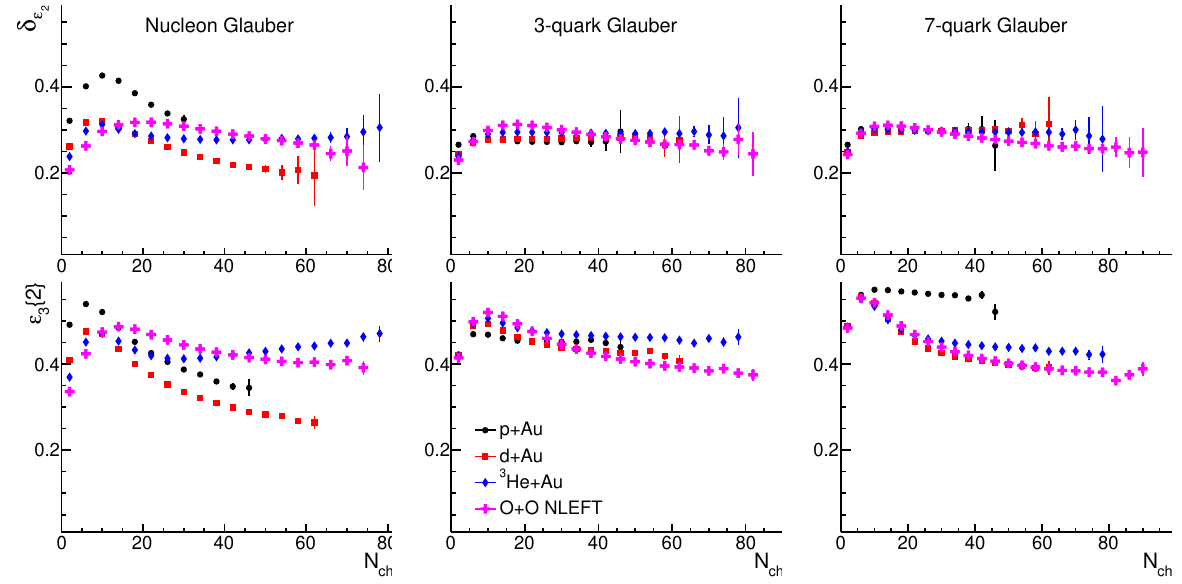}
\caption{Fluctuation-driven component of elliptic eccentricity $\delta_{\varepsilon_2}$ Eq.~\ref{eq:5} (top row) and $\varepsilon_{3}\{2\}$ (bottom row) for $p$+Au, $d$+Au, $^{3}$He+Au, and $^{16}$O+$^{16}$O collisions. Results are shown for nucleon Glauber (left), three-quark (middle), and seven-quark (right) Glauber models.}
\label{fig:3}
\end{figure*}

{\bf Results and discussion.} Figure~\ref{fig:1} shows the $p(\nch)$ distributions in various collision systems for the three Glauber models. We observe a crucial difference between symmetric and asymmetric systems: while the $p(\nch)$ for asymmetric collisions remain similar across three Glauber models, the O+O distributions broaden significantly when quark substructure is introduced. This occurs because tails of the distributions in asymmetric collisions are dominated by the configuration where the projectile $p/d/^3$He collides in the middle of the Au nucleus. Since the radius of $p/d/^3$He is much smaller than that of the Au nucleus, essentially all the sources within the projectile nucleus participate in the collision and the equivalent number of participating nucleons in the target Au nucleus differs modestly among the three Glauber models.

\begin{figure*}[htbp]
\centering
\includegraphics[width=0.85\linewidth]{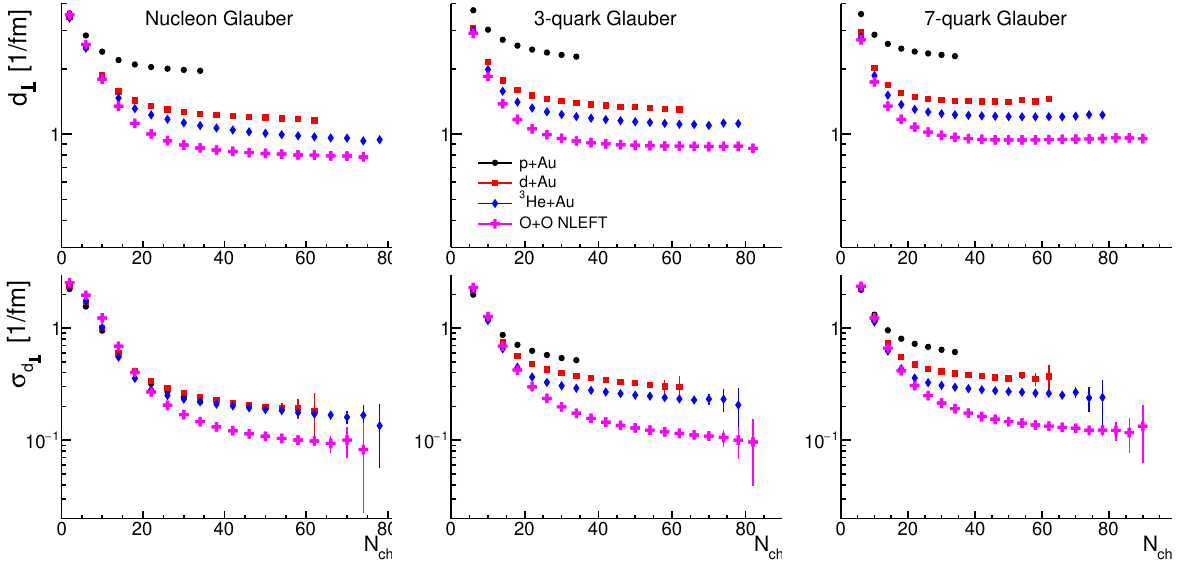}
\caption{Radial flow estimators as a function of $\nch$ in $p$+Au, $d$+Au, $^{3}$He+Au, and $^{16}$O+$^{16}$O collisions. Top row: inverse transverse size $d_{\perp}$ which is expected to correlate with event-wise mean transverse momentum $[\pT]$. Bottom row: variance of $d_{\perp}$, denoted as $\sigma_{d_{\perp}}$, which should correlate with fluctuations in $[\pT]$. Results are shown for the nucleon-level Glauber model (left), the three-quark Glauber model (middle ), and the seven-quark Glauber model (right).}
\label{fig:4}
\end{figure*}

On the other hand, in symmetric O+O collisions, many nucleons only have a fraction of their constituent quarks participating in the collision. This fraction is sensitive to the number of quark sources in each nucleon. Hence, having more sources per-nucleon significantly broadens the tails of the distribution in symmetric systems. This means that $p(\nch)$ in symmetric small collision systems are more sensitive to the subnucleonic fluctuations than the asymmetric collision systems. This provides a useful discriminator between nuclear and subnucleonic geometric effects.

Previous studies have shown that Glauber models considering nucleon substructure achieve a universal scaling of the multiplicity per-source i.e. $\lr{\nch}/\lr{N_s}$ as a function of centrality from $pp$, $p$+nucleus, small nucleus-nucleus, and large nucleus-nucleus collisions~\cite{PHENIX:2013ehw,PHENIX:2015tbb,Lacey:2016hqy,Zheng:2016nxx,Bozek:2016kpf}. The results in Fig.~\ref{fig:1} demonstrate that this scaling behavior is not necessary among $p/d/^3$He+Au systems, but is required when comparing them with the small symmetric collision systems, whose $p(\nch)$ distribution changes dramatically when the subnucleonic structure is considered. Previous PHENIX study has shown that the two-component Glauber model, where the particle production is driven by a $\npart$ component and a component proportional to number of binary collisions, can also accomplish similar level of agreement but for the wrong reason~\cite{PHENIX:2013ehw}. Our study suggests this inconsistency manifests already between asymmetric and small symmetric systems.

Figure~\ref{fig:2} demonstrates how subnucleonic fluctuations affect elliptic eccentricity in symmetric versus asymmetric systems. We shall first focus on $d$+Au, $^3$He+Au and O+O collisions. In the nucleon Glauber model, large system-dependent differences appear for $\varepsilon_2\{2\}$ and $\varepsilon_2\{4\}$. While their values agree at low-$\nch$ due to dominance of random fluctuations, a clear hierarchy appears at high $\nch$: $\varepsilon_2^{d\mathrm{Au}} > \varepsilon_2^{^3\mathrm{He+Au}} > \varepsilon_2^{\mathrm{O+O}}$ independent of Glauber model used. The pronounced elliptic shape of the deuteron contributes to the large $\varepsilon_2\{2\}$ and $\varepsilon_2\{4\}$ in $d$+Au collisions. The ratios of $\varepsilon_2\{4\}/\varepsilon_2\{2\}$ are similar among the asymmetric systems, which are all higher than O+O collisions. 

The ordering of $\varepsilon_2$ in $p$+Au relative to other systems varies strongly with the Glauber model used. $\varepsilon_2$ values are smallest in nucleon Glauber model, intermediate between $d$+Au and $^{3}$He+Au for the three-quark Glauber, and very close to values for $^{3}$He+Au for the seven-quark Glauber model. Given the large experimental uncertainties of $v_2^{p\mathrm{Au}}$ (about 30--40\%) associated with non-flow subtraction, the ordering of $\varepsilon_2$ among $p/d/^3$He+Au collisions are still compatible with the measured $v_2$ in the three systems~\cite{STAR:2023wmd}. On the other hand, $v_n$ have been measured very precisely in $p$+Pb collisions at the LHC~\cite{CMS:2015yux,ATLAS:2018ngv}. Hence, a comparison of $p$+Pb with O+O collisions at the LHC can unambiguously confirm whether nucleon Glauber or quark Glauber is preferred.

Figure~\ref{fig:3} shows the results of $\delta_{\varepsilon_2}$ and $\varepsilon_3\{2\}$, both of which reflect mainly the random fluctuations in the initial geometry. Sizable differences are observed among different collision systems in the nucleon Glauber model. However, adding quark substructure dramatically reduces these differences for $\varepsilon_3$, leading to similar values across all systems, consistent with the experimental observation of similar $v_3$ in $p/d/^3$He+Au collisions~\cite{STAR:2023wmd}. These behaviors suggest that $\varepsilon_3$ is dominated by fluctuations in both symmetric and asymmetric collisions. The only outlier is the $\varepsilon_3\{2\}$ in $p$+Au in the seven-quark Glauber model, which indicates a preference for the number of constituents to be less than seven.

If flow originates from geometry-driven hydrodynamic response, the patterns in Figs~\ref{fig:2}--\ref{fig:3} predict specific experimental signatures: If subnucleonic fluctuations are important, $v_2\{2\}$ and $v_2\{4\}$ should be largest in $d$+Au and smallest in O+O at the same $\nch$ at RHIC, while $v_3\{2\}$ should be similar across all systems. Furthermore, $v_2\{2\}$ and $v_2\{4\}$ in $p$+Au ($p$+Pb) collisions should be smaller than O+O collisions at the same $\nch$ at the RHIC(LHC), while $v_3\{2\}$ should be similar across all systems.

Figure~\ref{fig:4} presents predictions for radial flow observables through the inverse transverse size $d_{\perp}$ and its fluctuations $\sigma_{d_{\perp}} = \sqrt{\lr{(\delta d_{\perp})^2}}$. The $d_{\perp}$ is expected to be monotonically related to the $\lr{[\pT]}$, i.e. larger $d_{\perp}$ implies larger $\lr{[\pT]}$ and vice versa, although they are not proportional to each other. Therefore, one can only conclude that larger $\lr{[\pT]}$ is expected in smaller systems at the same $\nch$, but quantitative difference cannot be predicted by Glauber model. 

On the other hand, the linear response relation expected in hydrodynamics, $\delta[\pT] \propto \delta d_{\perp}$~\cite{Bozek:2012fw}, means that the variance of $d_{\perp}$ directly predicts the relative patterns of $\sigma_{[\pT]}=\sqrt{\lr{(\delta [\pT])^2}}$. Our study shows that the variance of $[\pT]$ should be significantly enhanced in $d$+Au collisions compared to O+O collisions at the same $\nch$, with the enhancement being larger when subnucleonic fluctuations are considered. In particular, the $\sigma_{d_{\perp}}$ values are similar among asymmetric collisions in nucleon Glauber model, while they show clear hierarchy in quark-Glauber model. These expectations provide an independent test of geometry-driven interpretation of the radial flow.

{\bf Summary and discussion.} We have demonstrated that comparing symmetric O+O collisions with asymmetric $p/d/^{3}$He+Au systems provides a unique approach to disentangle nuclear structure effects from subnucleonic fluctuations in small system flow. Our Glauber model calculations at both the nucleon and subnucleonic levels reveal distinct patterns in how these geometric sources affect different collision systems. Multiplicity distributions in O+O collisions show enhanced sensitivity to subnucleonic structure compared to asymmetric systems, serving as a clear discriminator.

Eccentricities exhibit system-specific hierarchies: $\varepsilon_2\{2\}$ and $\varepsilon_2\{4\}$ in $d$+Au, $^{3}$He+Au and O+O maintain the same ordering irrespective of subnucleonic fluctuations, with $\varepsilon_2\{2\}$ and $\varepsilon_2\{4\}$ in $d$+Au collisions remaining much larger than that in O+O collisions. In contrast, $\varepsilon_2\{2\}$ and $\varepsilon_2\{4\}$ in $p$+Au are smaller than O+O for nucleon Glauber model, but are larger for quark Glauber models. Furthermore, when subnucleonic fluctuations are included, $\varepsilon_3\{2\}$ and $\delta_{\varepsilon_2}$ become nearly universal. Finally, transverse size fluctuations reveal a robust ordering that provides an independent probe of the initial geometry through radial flow measurements.

These findings predict specific experimental signatures in flow observables that can distinguish a geometry-driven hydrodynamic response from alternative mechanisms like initial-state momentum anisotropy. The new $d$+Au and O+O data taken by the STAR experiment, as well as the new $p$+O and O+O runs at the LHC together with existing $pp$ and $p$+Pb data, will enable precise comparative measurements of $\nch$, $v_2$, $v_3$, and $[\pT]$ that will test our predicted orderings and help establish whether a geometry-driven final-state response dominates flow generation in small systems. 

Establishing the dominance of a geometry-driven final-state response is crucial to validate the application of the hydrodynamic framework to small systems. By constraining the relative contributions of nuclear and subnucleonic fluctuations, we can reduce systematic uncertainties in initial-condition modeling. This, in turn, could enable a more precise extraction of transport properties like shear viscosity and help determine if small systems produce the similar QGP observed in large collisions.

\begin{figure*}[htbp]
\centering
\includegraphics[width=0.7\linewidth]{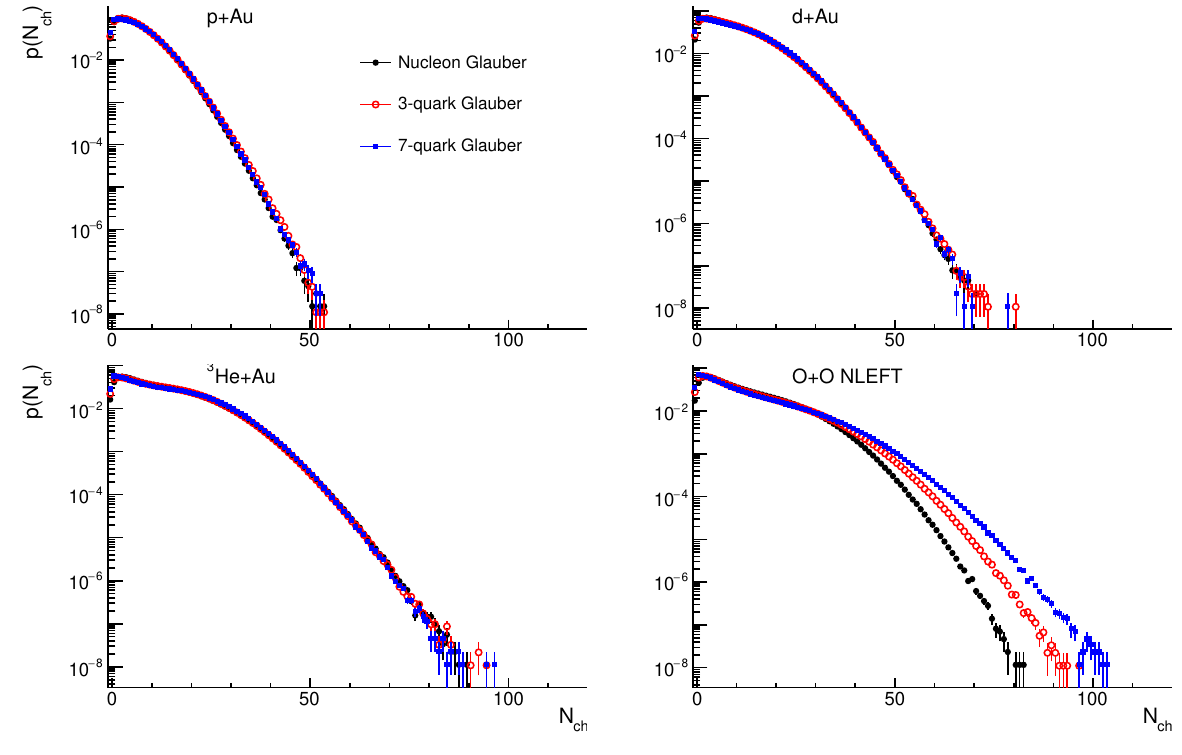}
\caption{The $\nch$ distribution for the three Glauber models in $p$+Au (top-left), $d$+Au (top-right), $^{3}$He+Au (bottom-left) and $^{16}$O+$^{16}$O (bottom-right) collisions at $\sqrt{s_{\mathrm{NN}}}$=200~GeV. The insensitivity of asymmetric collision systems to subnucleonic fluctuations, in contrast to symmetric O+O collisions, is clearly evident.}
\label{fig:5}
\end{figure*}
\begin{figure*}[htbp]
\centering
\includegraphics[width=0.85\linewidth]{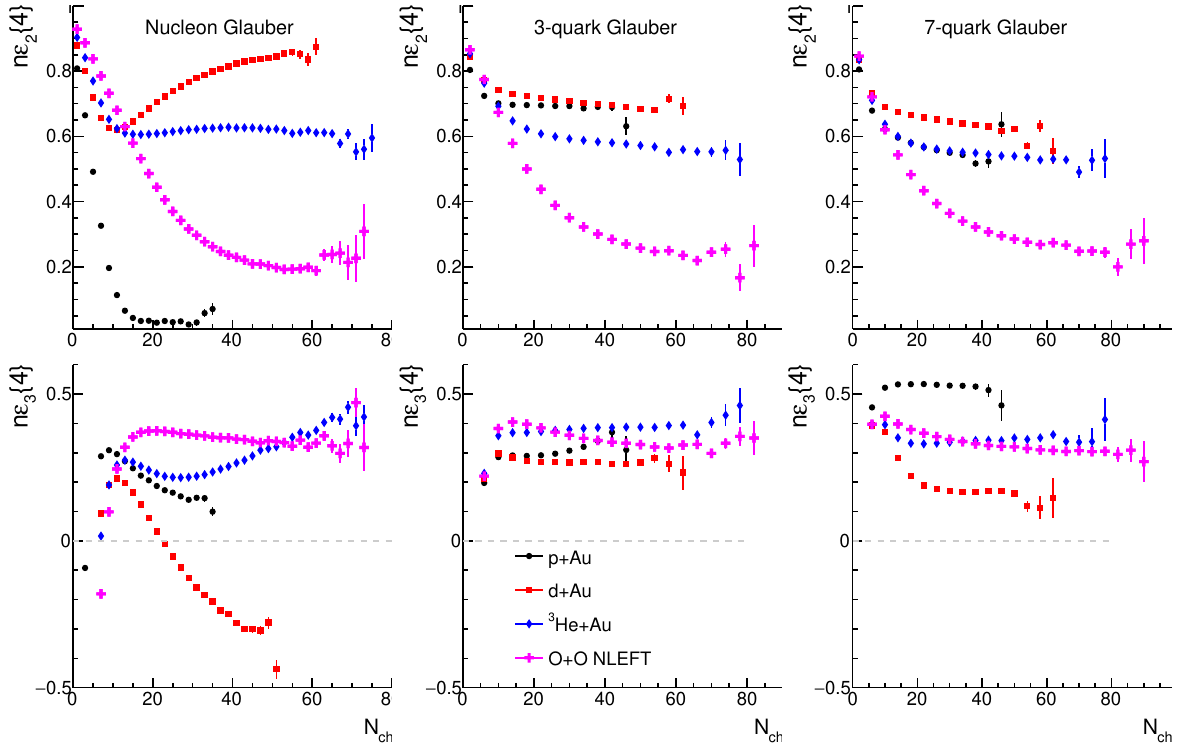}
\caption{Normalized four-particle cumulants as a function of $\nch$ in $p$+Au, $d$+Au, $^{3}$He+Au, and $^{16}$O+$^{16}$O collisions for elliptic eccentricity (top row) and triangular eccentricity (bottom row). Results are shown for nucleon Glauber (left), three-quark (middle), and seven-quark (right) Glauber models.}
\label{fig:6}
\end{figure*}
{\it Acknowledgement.} S. Huang and J. Jia are supported by the U.S. Department of Energy, Office of Science, Office of Nuclear Physics, under DOE Awards No. DE-SC0024602. C. Zhang is supported in part by the National Key Research and Development Program of China under Contract Nos. 2024YFA1612600, 2022YFA1604900, the National Natural Science Foundation of China (NSFC) under Contract Nos. 12025501, 12147101, 12205051, and Shanghai Pujiang Talents Program under Contract No. 24PJA009.

\section*{Appendix}
Figure~\ref{fig:5} shows the $p(\nch)$ distributions from Fig.~\ref{fig:1}, but reorganized in a different fashion to highlight the system-specific sensitivity to subnucleonic fluctuations. The insensitivity of asymmetric collision systems to the subnucleonic fluctuations, in contrast to symmetric O+O collisions, is more evident when displayed this way.

Figure~\ref{fig:6} compares the four-particle cumulants normalized by the two-particle cumulants
\begin{align}\label{eq:20}
\mathrm{nc}_n\{4\} \equiv \frac{\varepsilon_n\{4\}^4}{\varepsilon_n\{2\}^4} = 2-\frac{\lr{\varepsilon_n^4}}{\lr{\varepsilon_n^2}^2}\;,
\end{align}
for $n=2$ and 3. We see that the four-particle cumulants for third-order eccentricities show some significant splitting between different collision systems. In particular, the $\mathrm{nc}_3\{4\}$ in nucleon Glauber model is negative over a large range of $\nch$ in $^3$He+Au collisions, but remains positive in quark Glauber models. Furthermore, in seven-quark Glauber model, the $\mathrm{nc}_3\{4\}$ is larger in $p$+Au than other systems, which explains the anomalously larger $\varepsilon_3\{2\}$ values in $p$+Au in bottom right panel of Fig.~\ref{fig:3}.

\bibliography{smallsysfluc}{}
\bibliographystyle{apsrev4-1}

\end{document}